\documentclass[aps,prl,twocolumn]{revtex4-1}
\usepackage{amsmath}
\usepackage{amssymb}
\usepackage{graphicx}
\usepackage{braket}
\usepackage[colorlinks = true,linkcolor = red,urlcolor  = blue,citecolor = blue,anchorcolor = blue]{hyperref}

\newcommand{\phd}{\phantom{\dagger}}
\newcommand{\su}{\uparrow}
\newcommand{\sd}{\downarrow}

\newcommand{\nn}{\nonumber \\}

\newcommand{\expect}[1]{\left< #1 \right>}

\newcommand{\lbra}[1]{\langle {#1} |}
\newcommand{\rket}[1]{| {#1} \rangle}
\newcommand{\bracket}[2]{\langle {#1} | {#2} \rangle}

\begin{document}


\title{Proposal for measuring the parity anomaly in a topological superconductor ring}

\author{Chun-Xiao Liu}
\affiliation{Condensed Matter Theory Center and Joint Quantum Institute and Department of Physics, University of Maryland, College Park, Maryland 20742-4111, USA}

\author{William S. Cole}
\affiliation{Condensed Matter Theory Center and Joint Quantum Institute and Department of Physics, University of Maryland, College Park, Maryland 20742-4111, USA}

\author{Jay D. Sau}
\affiliation{Condensed Matter Theory Center and Joint Quantum Institute and Department of Physics, University of Maryland, College Park, Maryland 20742-4111, USA}

\date{\today}

\begin{abstract}
A topological superconductor ring is uniquely characterized by a switch in the ground state fermion number parity upon insertion of one superconducting flux quantum - a direct consequence of the topological ``parity anomaly.'' Despite the many other tantalizing signatures and applications of topological superconductors, this fundamental, defining property remains to be observed experimentally. Here we propose definitive detection of the fermion parity switch from the charging energy, temperature, and tunnel barrier dependence of the flux periodicity of two-terminal conductance of a floating superconductor ring. We extend the Ambegaokar-Eckern-Sch{\"o}n formalism for superconductors with a Coulomb charging energy to establish new explicit relationships between thermodynamic and transport properties of such a ring and the topological invariant of the superconductor. Crucially, we show that the topological contribution to the conductance oscillations can be isolated from Aharonov-Bohm oscillations of non-topological origin by their different dependence on the charging energy or barrier transparency.
\end{abstract}

\maketitle


Topological superconductors (TSC) are expected to support Majorana bound state excitations with non-Abelian statistics that might ultimately be harnessed for error-resistant quantum information processing~\cite{Nayak2008Non-Abelian, Alicea2012New, Leijnse2012Introduction, Beenakker2013Search, Stanescu2013Majorana, Jiang2013Non, DasSarma2015Majorana, Elliott2015Colloquium, Sato2016Majorana, Sato2017Topological, Aguado2017Majorana, Lutchyn2018Majorana}. Many simple, canonical examples of TSC have been theoretically formulated in one- and two-dimensional time-reversal-breaking superconductors (i.e., in class D)~\cite{Sau2010Generic, Lutchyn2010Majorana, Oreg2010Helical, Sau2010Non}, and several experiments now strongly suggest these have been realized in proximitized semiconductor nanowires among other systems~\cite{Mourik2012Signatures, Das2012Zero, Deng2012Anomalous,Churchill2013Superconductor, Finck2013Anomalous, Albrecht2016Exponential, Chen2017Experimental, Deng2016Majorana, Zhang2017Ballistic, Gul2018Ballistic, Nichele2017Scaling, Zhang2017Quantized}. However, despite the exciting progress that has been made, the experimental characterization of candidate TSC still admits some stubborn controversy. To date, most evidence comes from local probes, such as zero-bias anomalies in transport or excess zero-energy density of states, which indicate the presence of bound states~\cite{Mourik2012Signatures, Das2012Zero, Deng2012Anomalous,Churchill2013Superconductor, Finck2013Anomalous, Albrecht2016Exponential, Chen2017Experimental, Deng2016Majorana, Zhang2017Ballistic, Gul2018Ballistic, Nichele2017Scaling, Zhang2017Quantized}. The origin of controversy, though, is that \emph{any} bound state can always be decomposed, formally, into a pair of Majorana states so that even \emph{prima facie} dramatic transport phenomena such as the recently observed quantized zero-bias peak or an anomalous temperature scaling of a peak over a large temperature range can arise from a plausible ``quasi-Majorana'' situation where the probe predominantly couples to just one Majorana component of a bound state that, nevertheless, is not of topological origin, and does not have exponential-in-length insensitivity to local perturbations~\cite{Kells2012Near, Prada2012Transport, Liu2017Andreev, Chiu2017Conductance, Setiawan2017Electron, Liu2018Distinguishing,  Vuik2018Reproducing}.

Alternative methods to certify the existence of TSC are therefore desirable. The fractional Josephson effect (where the current-flux relationship has a $2\Phi_0$ periodicity, with $\Phi_0 = h/2e$ being the SC flux quantum) at a junction between topological superconductors has a particular appeal~\cite{Kitaev2001Unpaired, Kwon2004Fractional}. But in practice measuring this effect requires the junction to remain in a fixed fermion parity state and therefore must be observed at frequencies higher than the inverse parity lifetime~\cite{Lutchyn2010Majorana, Oreg2010Helical, Kitaev2001Unpaired, Kwon2004Fractional, Fu2009Josephson, Zocher2012Proposed}. In turn, ac measurement leads to complications such as Landau-Zener transitions~\cite{Sau2017Detecting}, which can yield a false positive in a topologically trivial state. The fractional Josephson effect however is merely an avatar of a more fundamental equilibrium topological property: the $Z_2$ ground state fermion parity of a TSC ring switches under the insertion of each SC flux quantum~\cite{Kitaev2001Unpaired, Sau2015Proposal, Cheng2015Fractional, Hell2018Distinguishing, Rubbert2016Detecting, Tripathi2016Fingerprints, Chiu2017Interference}.

\begin{figure}[t]
\begin{center}
\includegraphics[width=\linewidth]{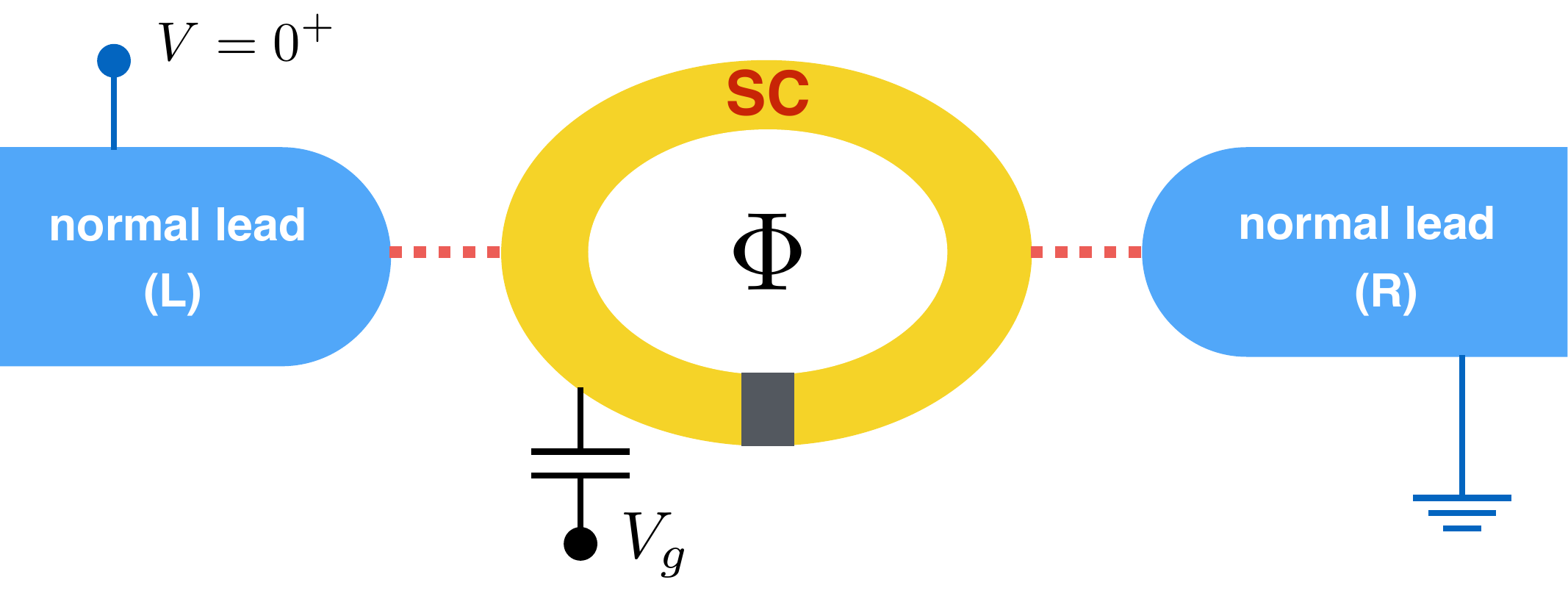}
\caption{Schematic of a proposed two-terminal transport experiment: a floating superconductor ring (yellow) is coupled to two normal metallic leads (blue). An infinitesimal bias voltage $V=0^+$ is applied across the leads. An insulating junction (gray) is present in the middle of the ring, and the enclosed magnetic flux $\Phi$ varies continuously along with the energy $\delta(\Phi)$ of an Andreev bound state at the junction.}
  \label{fig:schematic} 
  \end{center}
\end{figure}

In this Letter we describe a definitive transport measurement of this fermion parity switch. The essential principle is that a Coulomb charging energy $E_C$ promotes the parity anomaly into a genuine $2\Phi_0$ spectral periodicity \cite{Rubbert2016Detecting} (this is also related to its role in ``Majorana teleportation''~\cite{Fu2010Electron}), and this can be distinguished from conventional Aharonov-Bohm (AB) oscillations (which share the same periodicity) since the latter have no such dependence on $E_C$. To investigate this situation quantitatively, we have generalized the Ambegaokar-Eckern-Sch{\"o}n (AES) model to the case of a topological superconductor ring tunnel-coupling to external metallic leads. In this formalism we find that we can explicitly relate thermodynamic properties of the ring to the topological invariant, i.e., the ground state fermion parity.

The full Hamiltonian for the Coulomb blockaded normal-superconductor-normal (NSN) junction illustrated in Fig.~\ref{eq:fullHam} is
\begin{align}
&H = H_{\rm nw} + H_g + H_C + H_{\rm leads } + H_T, \nn
&H_{\rm nw} = \int \psi^{\dagger} \left( -\frac{\partial^2_x}{2m^*} - i \alpha \sigma_y \partial_x - \mu + V\sigma_z \right) \psi, \nn
&H_g = - g \int \psi^{\dagger}_{\su} \psi^{\dagger}_{\sd} \psi^{\phd}_{\sd} \psi^{\phd}_{\su}, \nn
&H_C = E_C \left( \int \psi^\dagger \psi - N_{\rm g} \right)^2, \nn
&H_{\rm leads } =\sum_{\rm L, R} \int \psi^{\dagger}_{\rm \alpha} \left( -\frac{\partial^2_x}{2m^*}  - \mu_{\rm \alpha}  \right) \psi_{\rm \alpha}, \nn
&H_T = -t \psi^{\dagger}_{\rm L}(\rm L) \psi(r_{\rm L}) -t \psi^{\dagger}_{\rm R}(\rm R) \psi(r_{\rm R})   + {\rm H. c.}.
\label{eq:fullHam}
\end{align}
$H_{\rm nw}$ is the semirealistic Majorana nanowire model \cite{Lutchyn2010Majorana, Oreg2010Helical} placed on a ring geometry, although we emphasize that the microscopic Hamiltonian for the SC ring will not be so essential in what follows. $H_g$ describes an attractive, local pairing interaction, and $H_C$ is the global charging energy relative to an induced charge $N_{\rm g}$. The Coulomb blockaded SC ring is weakly coupled to external leads on the left (L) and right (R) side, with typical lead ($H_{\rm leads}$) and coupling ($H_T$) Hamiltonians.


\begin{figure}[t]
\begin{center}
\includegraphics[width=\linewidth]{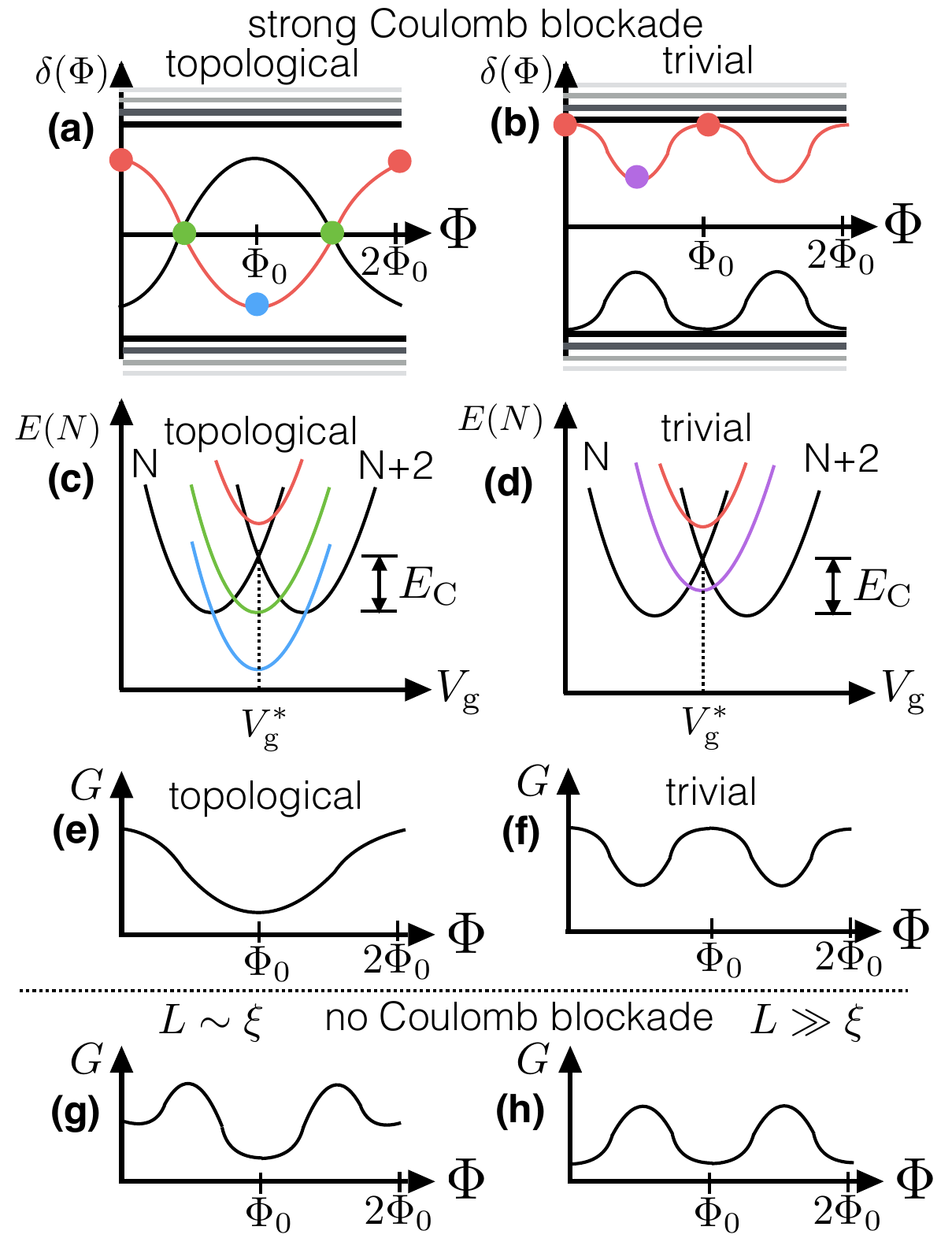}
\caption{Large and no charging energy limits. (a) and (b) are energy spectra for the bound states at the junction of a topological or trivial SC ring. (c) and (d) are the evolution of the energy of the $N+1$ charge state as a function of the magnetic flux ($\Phi$). Each color line corresponds to the color point in the spectra in (a) and (b), and $V^*_{\text{g}}$ represents the resonance point at which the $N$ and $N+2$ charge states are degenerate. (e) and (f) are conductance for the NSN junction with strong Coulomb blockade. Topological (trivial) SC shows $2\Phi_0$ ($\Phi_0$) periodicity. (g) and (h) are conductance for the NSN junction with no Coulomb blockade. Here, the short (long) SC ring shows $2\Phi_0$ ($\Phi_0$) periodicity in conductance, regardless of SC being topological or trivial.}
\label{fig:strong_weak} 
\end{center}
\end{figure}

\emph{Large $E_C$ and $E_C=0$ limits} ---
The conductance in each of these cases can be understood qualitatively as shown in Fig.~\ref{fig:strong_weak}. We first consider large $E_C$ \cite{Albrecht2016Exponential, vanHeck2016Conductance}, and an idealized low-energy limit (not essential for later) of the microscopic model: a single subgap state bound to the junction with energy $\delta(\Phi)$. The Bogoliubov-de Gennes (BdG) energy spectrum is $\Phi_0$-periodic in both the trivial and topological cases, but the latter has a parity switch and the former does not [Fig.~\ref{fig:strong_weak}(a) and (b)]. In conventional NSN Coulomb blockade, sharp zero-bias conductance peaks occur when the induced charge $e N_{\rm g}$ is tuned to degeneracy between charge states of the island separated by $2e$, and Andreev reflection (transferring a charge-$2e$ Cooper pair to the island) is enabled. If for any flux $\Phi$ the energy of the subgap state is \emph{lower} than the charging energy of those two degenerate states [Fig.~\ref{fig:strong_weak}(c) and (d)], then Andreev conductance is suppressed as the island relaxes to the new nondegenerate ground state at this formerly resonant value of induced charge ($V_{\text{g}} =V^*_{\text{g}}$). Fig.~\ref{fig:strong_weak}(e) and \ref{fig:strong_weak}(f) show schematically that even though the BdG energy spectrum $\delta(\Phi)$ is $\Phi_0$-periodic, the spectrum of $H_{\text{C}}$ in the presence of this subgap state, and the corresponding conductance, need not be; in the topological case the conductance period is doubled.

By setting $E_C$ to zero, on the other hand, there is no Coulomb blockade of Andreev processes. The conductance for the two-terminal junction with a floating superconductor is \cite{Takane1992Conductance, Anantram1996Current,Ulrich2015Floating}
\begin{align}
G = \frac{e^2}{h} \frac{  g_{\text{LL}}g_{\text{RR}} -  g_{\text{LR}}g_{\text{LR}} }{ g_{\text{LL}} + g_{\text{RR}} -  g_{\text{LR}} - g_{\text{LR}} }.
\label{eq:Gsmatrix}
\end{align}
Here, $g_{\text{LL}}$ ($g_{\text{RR}}$) is the dimensionless local conductance for the left (right) lead, while $g_{\text{LR}}$ ($g_{\text{RL}}$) is the dimensionless conductance from the right (left) lead to the left (right) lead. In the short ring limit, the conductance for the NSN junction is always $2\Phi_0$-periodic, regardless of the SC ring being in the topological or trivial phase [Fig.~\ref{fig:strong_weak}(g)], as single-quasiparticle interference processes contribute to all $g_{\alpha \beta}$. In the long ring limit, single-quasiparticle interference is generally suppressed. As the transport coefficients are associated with the BdG energy spectrum only, which is $\Phi_0$-periodic, conductance for a long SC ring is likewise $\Phi_0$-periodic [Fig.~\ref{fig:strong_weak}(h)]. Thus, the conductance of the NSN junction absent charging energy cannot distinguish between a topological or a trivial origin (i.e., arising from a short ring, or low-energy states due to disorder or order parameter fluctuations) of the doubled periodicity.


\emph{Generalized AES model} ---
To study the properties of the superconductor ring of Fig.~\ref{fig:schematic} beyond the qualitative limits of the previous section, we analyze Eq.~\eqref{eq:fullHam} in an imaginary-time path integral formulation in the spirit of AES~\cite{Ambegaokar1982Quantum}. Details are in \cite{Supp}, and we outline the procedure here. The partition function of the system can be written as $Z = \mbox{Tr} \, e^{-\beta H} \equiv \int \mathcal{D}\bar{\psi}\mathcal{D}\psi \, e^{-S[\bar{\psi}, \psi]}$. The infinite leads are replaced by self-energies $\Sigma_{\alpha}(\tau)= \frac{-\Gamma_{\alpha}/\beta}{\sin(\pi \tau/\beta)}$. For the quartic terms $H_g$ and $H_C$, we perform the standard Hubbard-Stratonovich transformations to replace them with an imaginary-time-varying SC pairing potential $\Delta(x, \tau) e^{i \phi(x, \tau)}$ and an electrostatic potential $V(\tau)$ tracking total charge fluctuations, and to make the problem tractable, we focus on fluctuations around the saddle point with constant $\Delta(x, \tau) = \Delta_0$, valid for $T \ll \Delta_0$, and $\phi(x, \tau) = \phi(\tau)$. Note that the effect of the magnetic flux threading through the ring is now absorbed in $H_{\rm nw}$.

To eliminate the $\phi$ dependence from the effective fermion Hamiltonian, we make a gauge transformation to the fermion fields $\psi(x,\tau) \rightarrow \psi'= \exp(i\phi/2) \psi$. However, we observe that this results in an atypical boundary condition for fermions: $\psi'(\beta) = -\exp(i\pi W) \psi'(0)$, where $W = \frac{1}{2\pi}\int_0^\beta d\tau (\partial_\tau \phi) = \frac{1}{2\pi}[\phi(\beta) - \phi(0)]$ is the integer winding number of the phase field. In other words, $\psi'$ is antiperiodic or periodic in $\beta$ depending on whether the winding number $W$ is even or odd. This gauge transformation further results in an effective chemical potential variation $\delta\mu = i(\partial_\tau \phi/2 + V)$. Fixing $\delta\mu = 0$, in the same saddle point approximation, results in the Josephson relation $V(\tau) = -\partial_\tau \phi / 2$ locking charge and phase fluctuations, after which we can finally integrate out the quadratic fermion fields.

Following these mostly standard manipulations, we begin to approach one of our central results: the only remaining degree of freedom in the effective action is the phase variable, and the partition function can be decomposed into discrete topological sectors indexed by $W$. Formally, then, the partition function is written as
\begin{equation}
Z = \sum_{W} Z_W = \sum_{W} Z^{\rm BdG}_{W} \int_W \mathcal{D}\phi e^{-S_W[\phi]},
\end{equation}
where, first, $Z_W^{\rm BdG}$ results from integrating out the $\psi$ fields subject to the boundary condition stated above. Originating from the correspondence between boundary condition (and thus Fourier expansion in boson or fermion Matsubara frequencies) and winding number, we obtain that the topological invariant enters the partition function explicitly, depending on the parity of $W$,
\begin{align}
Z_{{\rm even} \; W}^{\rm BdG} &=
\prod_{\varepsilon > 0} 2 \mbox{cosh} \left( \frac{\beta \varepsilon}{2} \right) \\
Z_{{\rm odd} \; W}^{\rm BdG} &= \left( \mbox{sgn} \; \mbox{Pf} \; H_{\rm BdG} \right)
\prod_{\varepsilon > 0} 2 \mbox{sinh} \left( \frac{\beta \varepsilon}{2} \right)
\end{align}
where $H_{\rm BdG}$ is the mean-field quadratic Hamiltonian appearing in the action after the Hubbard-Stratonovich transformation, written in the Majorana basis, and $\varepsilon$ are its positive eigenvalues. It is useful at this point to note that there is \emph{no} direct correspondence between winding number and the parity of occupied quasiparticle states, so this decomposition is conceptually distinct from prior works where the partition function is written as a sum of odd and even quasiparticle occupation parity sectors. We do recover, however, an equivalent partition function (see e.g. \cite{Beenakker2013Anomaly}) in the appropriate $E_C = 0$ limit.

Next, the remaining effective action for the phase, $S_W[\phi] = S^0_W[\phi] + S^{\rm leads}_W[\phi]$, consists of the familiar ``particle on a ring'' (n.b., in imaginary time, rather than real space) with a topological term proportional to the induced charge
\begin{equation}
S^0_W[\phi] = \int d\tau \frac{(\partial_\tau \phi)^2}{4E_C} -i\pi N_{\rm g} W
\end{equation}
and a dissipative contribution arising from the tunnel-coupling to the external leads
\begin{align}
S^{\rm leads}_W[\phi] &= - \frac{1}{2}\mbox{Tr} \log ( 1 - G_{\rm SC} \Sigma ) \nn
 &\simeq g_0 \int \frac{d\tau_1 d\tau_2}{\beta^2} \frac{1-\cos[ \phi(\tau_1) - \phi(\tau_2) ]}{\sin^2[ \pi (\tau_1-\tau_2)/\beta ]},
\end{align}
where $g_0 = (g_{\rm LL} + g_{\rm RR})/2$ is the dimensionless local conductance averaged over left and right leads. We have assumed that the tunneling strength between the SC island and the leads $\Gamma_{\alpha}$ is weak, and that the Green's function of the SC island $G_{\rm SC}(x,\tau)$ is local in both space and time~\cite{Supp}.

Summarizing so far, we have derived an effective action in the spirit of the AES model, and in doing so we made manifest the relationship between the imaginary-time winding number of the effective phase degree of freedom and the ground state parity of the superconductor, expressed as Kitaev's topological invariant~\cite{Kitaev2001Unpaired}. The charging energy controls the relative contribution of different winding number sectors to the full partition function. Therefore, flux period doubling arising from the topological parity switch has explicit Coulomb dependence, whereas any conventional Aharanov-Bohm periodicity appears already in $Z_0$ with no dependence at all on $E_C$. In other words, topological and non-topological period doubling can be disentangled even in a device where the latter is present.

\emph{Measurement} ---
Like the partition function itself, any equilibrium observable can be expanded in $W$ sectors and evaluated independently in each. To facilitate this, for each $W$ we can take $\phi(\tau) = 2 \pi W \tau / \beta + \delta\phi(\tau)$, where $\delta\phi(0) = \delta\phi(\beta)$ so all the winding is contained in the first part. With this substitution
\begin{equation}
S_W^{0}[\phi] = \frac{\pi^2 W^2}{\beta E_C}   -i \pi N_{\rm g} W + \int d \tau \frac{(\partial_\tau \delta\phi)^2}{4 E_C}
\end{equation}
which heavily suppresses large winding number contributions for intermediate temperatures $E_C \lesssim T \ll \Delta_0$. Continuing in this regime, we also obtain to zeroth order in $\delta\phi$ that $S^{\rm leads}_W= 2 g_0 |W|$, so that, approximately,
\begin{align}
Z_{\pm 1} / Z_0 \approx  \left( \mbox{sgn} \; \mbox{Pf} \; H_{\rm BdG} \right)
\prod_{\varepsilon > 0} 2 \mbox{tanh} \left( \frac{\beta \varepsilon}{2} \right) \times \nn
\exp \left( \pm i \pi N_{\rm g} \right) \exp \left( -\frac{\pi^2}{\beta E_C} - 2 g_0 \right)
\label{eq:Z1overZ0}
\end{align}
and so any ground-state parity dependence can be equivalently eliminated by (i) lowering $E_C$, (ii) increasing temperature, or (iii) increasing the barrier transparency and therefore $g_0$, all of which tend to favor a pinned phase $\phi$.

To quadratic order in $\delta\phi$ we next calculate the zero-bias conductance~\cite{Bascones2000Nonequilibrium} of the device in Fig.~\ref{fig:schematic} as
\begin{align}
G &\approx g_0\expect{e^{i\phi(\beta/2)-i\phi(0)}}\nn
&\approx G_0 + (G_1 - G_0) \frac{Z_1}{Z_0} + (G_{-1} - G_0) \frac{Z_{-1}}{Z_0}
\label{eq:Gtot}
\end{align}
up to exponentially small corrections in $g_0$ and $(\beta E_C)^{-1}$. Equations~\eqref{eq:Z1overZ0} and \eqref{eq:Gtot} illuminate the behavior of the weakly Coulomb blockaded SC ring. As the ground state parity of the SC ring is now contained in the ratio $Z_{\pm 1}/Z_0$, when the charging energy $E_C$ goes to zero, this ratio is exponentially suppressed. Correspondingly, the conductance without Coulomb blockade cannot give any information about the ground state parity of the SC ring. Instead, the conductance without Coulomb blockade is fixed by the BdG spectrum and quasiparticle wavefunctions of the isolated ring. $G_0$ in Eq.~\eqref{eq:Gtot} will be $\Phi_0$ ($2\Phi_0$)-periodic, when the length of the ring is long (short) compared to the coherence length and the AB effect is suppressed (prominent). This asymptotic behavior based on our partition function calculation is consistent with the discussion following Eq.~\eqref{eq:Gsmatrix}.

\begin{figure}[t]
\begin{center}
\includegraphics[width=\linewidth]{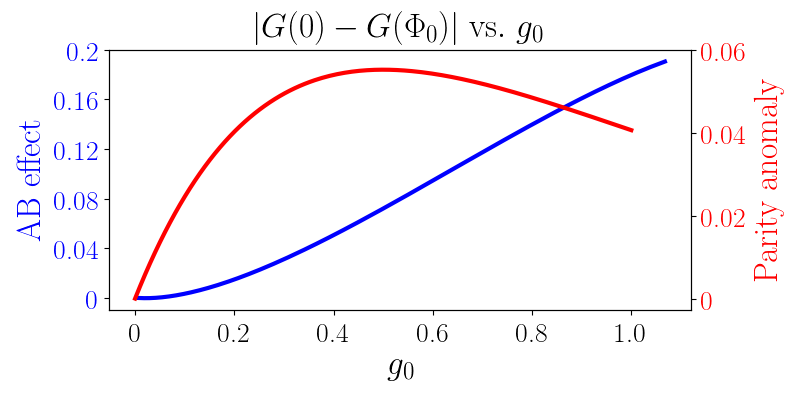}
\caption{Conductance difference $G(0) - G(\Phi_0)$ as a function of the barrier transparency characterized by $g_0$. For a short SC ring (no matter topologically trivial or non-trivial) with no charging energy, the conductance difference increases monotonically with barrier transparency (blue line). By contrast, for a TSC ring with charging energy, although increasing at low barrier transparency, the conductance difference will eventually decrease with barrier transparency when $g_0$ goes beyond some critical value $\sim 0.5$ (red line). }
\label{fig:dG_vs_g0} 
\end{center}
\end{figure}

In Fig.~\ref{fig:dG_vs_g0}, we plot the conductance difference after flux insertion, $G(0) - G(\Phi_0)$, as a function of the lead-SC interface conductance $g_0$ (since we are not in the strong Coulomb blockade limit, this is calculated at $N_{\rm g} = 0$). A nonzero value of this conductance difference is a direct indication of $2\Phi_0$ periodicity, and $g_0$ is realistically tunable by a tunnel barrier. We consider first a short SC ring (it does not matter if the SC is topologically trivial or nontrivial) without Coulomb blockade and calculate the conductance using Eq.~\eqref{eq:Gsmatrix}. The resulting conductance difference is shown by the blue line in Fig.~\ref{fig:dG_vs_g0}. The signal of trivial AB-induced $2\Phi_0$ periodicity is monotonically increasing with the junction conductance $g_0$. For comparison, for the NSN junction with finite charging energy and topological SC ring, the conductance difference calculated by Eq.~\eqref{eq:Gtot} is shown as the red line in Fig.~\ref{fig:dG_vs_g0}. Note that although the conductance difference initially increases with small $g_0$, beyond some critical value $g_0 \simeq 0.5$, the signal of $2\Phi_0$ periodicity \emph{decreases} with the conductance. This is because the large tunnel transparency effectively renormalizes the charging energy $E_C$ to a smaller value and thus suppresses the parity anomaly-induced $2\Phi_0$ periodicity. Practically, if a decrease of $2\Phi_0$ periodicity with increasing tunnel conductance is observed experimentally, it would indicate a topologically nontrivial Coulomb blockaded superconductor. In reality, the signal may arise from both AB effect and parity anomaly, and the relative strength of the two is unknown \emph{a priori}. We note, however, that the AB contribution can also be systematically suppressed by decreasing the junction transparency in the ring, and the parity anomaly contribution can be systematically increased by lowering the temperature.

\emph{Discussion} --- Our proposal relates the two-terminal zero-bias conductance of the device in Fig.~\ref{fig:schematic} to the fundamental equilibrium parity anomaly of the \emph{bulk} topological superconductor, independent of the presence of Majorana modes or non-Abelian statistics. Accordingly, accidental, near zero-energy ABS cannot alter this topological property of the ring to produce a false positive signature. In terms of feasibility, all the ingredients for this proposal are separately in place in previous experiments: (1) nanowire rings or ``hashtags'' demonstrating conventional Aharonov-Bohm oscilliations in the absence of superconductivity \cite{Gazibegovic2017Epitaxy}, (2) two-terminal proximity-SC islands, where Coulomb blockade can be tuned via the transparency of one of the barriers, and (3) robust Zeeman-tuned parity switches in Coulomb blockaded class D islands \cite{Shen2018Parity}, indicating the absence of any substantial density of subgap states. It remains now to combine these ingredients. Although long parity lifetimes and protection from nonequilibrium quasiparticles will eventually be necessary for quantum information applications, they are not requirements for the definitive transport measurement of the parity anomaly we have discussed. Finally, from a theorist's perspective, we expect our generalization of the AES model, incorporating the mean-field topological invariant $(\mbox{sgn} \; \mbox{Pf} \; H_{\rm BdG})$ and thereby exploring its beyond-mean-field consequences, could motivate related generalizations for floating topological superconductors and quantum dots in other symmetry classes.


We thank C.-K. Chiu for many helpful discussions. C.-X.L. was supported by Microsoft Station Q, and acknowledges the hospitality of the Kavli Institute of Theoretical Sciences at University of Chinese Academy of Sciences. W.S.C. was supported by LPS-MPO-CMTC. J.S. acknowledges support from the Alfred P. Sloan foundation and the National Science Foundation NSF DMR-1555135 (CAREER). 

\bibliography{BibMajorana}

\onecolumngrid

\section{Supplementary material }

\subsection{ I. Conductance with no Coulomb blockade }
The conductance for an NSN junction with no charging energy can be calculated using the generalized Landauer-B\"uttiker formalism~\cite{Takane1992Conductance, Anantram1996Current}:
\begin{align}
I_{1} &= \frac{e^2}{h} \Big[ g_{11} (V_{1} - V_{\text{SC}}) + g_{12} (V_{2} - V_{\text{SC}}) \Big], \nn
I_{2} &= \frac{e^2}{h} \Big[ g_{21} (V_{2} - V_{\text{SC}}) + g_{22} (V_{2} - V_{\text{SC}}) \Big]
\end{align}
where $I_{i}$ is the current flowing inside the $i$-th lead towards the device, $V_i$ ($V_{\text{SC}}$) is the voltage for the $i$-th lead (floating superconductor). The microscopic expression for $g_{ij}$ in terms of the transmission amplitude is
\begin{align}
&g_{ij}= \delta_{ij}N - t_{ij}= \sum_{\alpha, \beta =\su, \sd } \Big( \delta_{ij}\delta_{\alpha \beta} - t_{ij,\alpha \beta} \Big) , \nn
&t_{ij,\alpha \beta} = \int dE \Big(-f'_j(E) \Big)  \Big[ |s^{\text{ee}}_{ij,\alpha \beta}(E)|^2 - |s^{\text{he}}_{ij,\alpha \beta}(E)|^2 \Big],
\end{align}
where $N$ is the total number of spinful channels in the lead, $s^{\text{ee}}_{ij,\alpha \beta}$($s^{\text{he}}_{ij,\alpha \beta}$) is the normal (Andreev) transmission amplitude for an incoming electron with spin-$\beta$ in the $j$-th lead to become an outgoing electron (hole) with spin-$\alpha$ in the $i$-th lead. As the superconductor is floating, the current conservation ($I_1+I_2=0$) gives
\begin{align}
G = \frac{I_1}{V_1 - V_2}=\frac{e^2}{h} \frac{g_{11}g_{22} -g_{12}g_{21} }{g_{11}+g_{22} +g_{12}+g_{21} } = \frac{e^2}{h} \frac{g_{11}g_{22} -t_{12}t_{21} }{g_{11}+g_{22} -t_{12}-t_{21} }.
\end{align}
Physically, $g_{ii}$ is the local dimensionless conductance through the interface between the $i$-th lead and the SC island, and $t_{ij}$ is the dimensionless conductance from the $j$-th lead to the $i$-th lead.


\subsection{II. Self-energy from the normal-metal lead }
\label{sec:self-energy}
When a system is coupled to a metallic lead, the influence from the lead can be summarized in a term called self-energy for the system. This self-energy term shows up in a clear way in the path integral formalism. The partition function for the coupling between the system and the metallic lead is
\begin{align} 
&Z = \int \mathcal{D} \overline{\psi}_{\alpha} \mathcal{D} \psi_{\alpha} \exp\{ - (S_{\alpha} + S_{\rm T})  \} \nn
&S_{\alpha}  = \int d\tau dx  \overline{\psi}_{\alpha}(\tau,x)  ( \partial_{\tau} + h_{\alpha}  ) \psi_{\alpha} (\tau,x) \nn
&S_{\rm T} = \int d\tau \Big[ -t\overline{\psi}_{\alpha}(\tau, R ) \psi_0(\tau, x_0) -t^* \overline{\psi}( \tau, x_0) \psi_{\alpha} ( \tau, R) \Big],
\end{align}
where $S_{\alpha}$ is the action of the lead, and $S_{\rm T}$ describes the  tunneling of an electron between $x_0$ on the island and $R$ in the lead. After completing the square for Grassmann fields of the lead and integrating them out, we get
\begin{align}
Z &= \int \mathcal{D} \overline{\psi}_{\alpha} \mathcal{D} \psi_{\alpha} \exp \Big\{ \int  \Big( \overline{\psi}_{\alpha} + \int \overline{\psi}_0t^*G  \Big) G^{-1} \Big( {\psi}_{\alpha} + \int Gt   {\psi}_0 \Big) - \int \overline{\psi}_0 t^* G  t{\psi}_0 \Big\} \nn
& \propto \exp \Big\{ - \int d\tau_1 d\tau_2 \overline{\psi}(\tau_2, x_0) t^* G(R, \tau_2 - \tau_1)  t {\psi}(\tau_1, x_0)  \Big\}
\end{align}
where $G=-\partial_{\tau} - h_{\alpha}$ is the Green's function for the isolated lead, and the self-energy from the lead defined as 
\begin{align}
\Sigma(R, \tau_2-\tau_1) = t^* G(R, \tau_2- \tau_1)t.
\end{align}
The self-energy describes the process during which an electron tunnels from the island into the lead, stays inside, and finally hops back to the island. Although the tunneling process is local in space, it is nonlocal in time. The self-energy term has a closed form in the frequency representation:
\begin{align}
\Sigma(R, i\omega_n) &= t^2 G(R, i\omega_n) =t^2 \lbra{R} (i\omega_n - h_{\alpha})^{-1} \rket{R} = t^2\sum_{k} | \bracket{R}{\epsilon_k} |^2 \frac{1}{i\omega_n - \epsilon_k} \nn
& \simeq t^2 d(\epsilon_F) | \bracket{R}{\epsilon_F} |^2 \int^{+\infty}_{-\infty} d \epsilon  \frac{1}{i\omega_n - \epsilon} = -i\Gamma {\rm sgn} (\omega_n),
\end{align}
where $i\omega_n = (2n+1)\pi/\beta$ is the Matsubara frequency for lead electrons and $\Gamma =\pi t^2 d(\epsilon_F)|\bracket{R}{\epsilon_F}|^2$ is the tunneling strength proportional to the local density of states at fermi surface at the lead end. To obtain the self-energy in the temporal representation, one needs to first  introduce an exponential ultra-violet suppression factor $e^{-|\omega_n|/\Lambda}$ and set $\Lambda$ to infinity in the end of the calculation:
\begin{align}
\Sigma(\tau) &= -i\Gamma \frac{1}{\beta}\sum_{\omega_n} {\rm sgn} (\omega_n) e^{-i\omega_n\tau}e^{-|\omega_n|/\Lambda} \nn
&=-i\Gamma \frac{1}{\beta} \sum_{\omega_n>0} {\rm sgn} (\omega_n) \Big[ e^{(-i\tau - \Lambda^{-1}) \omega_n} - e^{(i\tau - \Lambda^{-1}) \omega_n} \Big]\nn
&= -i\Gamma \frac{1}{\beta} \Big( \frac{e^{-i\tau\pi/\beta}}{1-e^{-i\tau2\pi/\beta}} - \frac{e^{i\tau\pi/\beta}}{1-e^{i\tau2\pi/\beta}} \Big) \nn
&= \frac{-\Gamma/\beta}{\sin(\pi \tau/\beta)}.
\end{align}


\subsection{III. Partition function for the NSN junction with charging energy}

The action for the Coulomb-blockaded NSN junction in the imaginary path integral formalism is
\begin{align}
&Z = \int \mathcal{D}\overline{\psi}\mathcal{D}{\psi}e^{-S},\nn
&S = S_0 + S_{g} + S_{C} + S_{\rm{lead}}, \nn
&S_0 = \sum_{\alpha, \beta = \su \sd} \int d\tau dx \overline{\psi}_{\alpha} \Big[ \partial_{\tau}\delta_{\alpha \beta} + h_{\rm{nw}}^{\alpha \beta}(x) - \mu\delta_{\alpha \beta} \Big]{\psi}_{\beta}(\tau, x), \nn
&S_{g} = -g \int d\tau dx \overline{\psi}_{\downarrow}\overline{\psi}_{\uparrow}{\psi}_{\uparrow}{\psi}_{\downarrow}(\tau, x), \nn
&S_{C} = E_C  \int^{\beta}_0 d\tau \Big[ N(\tau) - N_{\rm g} \Big]^2, \nn
&S_{\rm{lead}} = \sum_{\alpha  = \su \sd} \sum_{a=L,R} \int d\tau_2 d\tau_1 \overline{\psi}_{\alpha}(\tau_2, x_a ) \Sigma_a( \tau_2-\tau_1) {\psi}_{\alpha}( \tau_1, x_a),
\end{align}
where $S_0$ is the action for the semiconductor nanowire including the insulating junction and the magnetic flux dependence, $S_g$ is the point-like attraction between electrons, $S_C$ is the Coulomb energy for the island with $ N_{\rm g}$ being the induced charge, and $S_{\rm lead}$ is the self-energy from the two normal-metal leads, as derived in Sec.~\ref{sec:self-energy}. $x_a$ is the endpoint on the nanowire coupling to the lead. Next, we perform the Hubbard-Stratonovich transformation on $S_g$ and $S_C$, so that the partition function now becomes
\begin{align}
&Z = \int \mathcal{D}\overline{\psi}\mathcal{D}{\psi}\mathcal{D}\overline{\Delta}\mathcal{D}{\Delta}\mathcal{V}e^{-S},\nn
&S = S_0 + S_{\rm lead} + \int d\tau dx \Big( \Delta(\tau, x) {\psi}_{\uparrow}{\psi}_{\downarrow} + \Delta^*(\tau, x)\overline{\psi}_{\downarrow}\overline{\psi}_{\uparrow} \Big) - i\int d\tau V(\tau)\Big( N(\tau) - N_{\rm g} \Big) + \int d\tau dx \frac{|\Delta(\tau, x)|^2}{g} + \int d\tau \frac{V(\tau)^2}{E_C},
\end{align}
where the complex auxiliary field $\Delta(\tau, x)$ is the superconducting pairing potential, and the real auxiliary field $V(\tau)$ is the electrostatic potential tracking the charge fluctuations. To make the problem tractable, we only consider the phase fluctuations around the saddle point of constant pairing amplitude: $\Delta(\tau, x) = |\Delta(\tau, x)| e^{i \phi(\tau, x)} \approx \Delta_0 e^{i\phi(\tau)}$. Next, we make a gauge transformation to the electron field:
\begin{align}
&\psi_{\alpha}(\tau, x) \to \psi'_{\alpha}(\tau, x)=e^{i\phi(\tau)/2} \psi_{\alpha}(\tau, x), \nn
&\overline{\psi}_{\alpha}(\tau, x) \to \overline{\psi}'_{\alpha}(\tau, x)=e^{-i\phi(\tau)/2} \overline{\psi}_{\alpha}(\tau, x),
\end{align}
so that the partition function in terms of the new electron field is
\begin{align}
Z &= \int \mathcal{D}\overline{\psi}' \mathcal{D}{\psi}'\mathcal{D}{\phi}\mathcal{D}\mathcal{V}e^{-S},\nn
S &=  \sum_{\alpha, \beta = \su \sd} \int d\tau dx \overline{\psi}'_{\alpha} \Big[ \Big( \partial_{\tau} - i \partial_{\tau} \phi/2 - iV(\tau)  -\mu \Big) \delta_{\alpha \beta} + h_{\rm{nw}}^{\alpha \beta}(x)  \Big]{\psi}'_{\beta}(\tau, x) + \Delta_0 \int d\tau dx \Big(  {\psi}'_{\uparrow}{\psi}'_{\downarrow} + \overline{\psi}'_{\downarrow}\overline{\psi}'_{\uparrow} \Big) \nn
&+ \sum_{\alpha  = \su \sd} \sum_{a=L,R} \int d\tau_2 d\tau_1 \overline{\psi}'_{\alpha}(\tau_2, x_a ) \Sigma'_a(\tau_2, \tau_1) {\psi}'_{\alpha}( \tau_1, x_a)  + iN_{\rm g} \int d\tau V(\tau) + \int d\tau \frac{V(\tau)^2}{E_C},
\end{align}
where $\Sigma'_a(\tau_2, \tau_1) = \Sigma_a( \tau_2-\tau_1)e^{i [\phi(\tau_2) - \phi(\tau_1)]}$. Note that this transformation results in an atypical boundary condition for the fermions
\begin{align}
&\psi'(\beta) = -e^{i\pi W}\psi'(0), \nn
&W = \frac{1}{2\pi} \int^{\beta}_0 \partial_{\tau}\phi = \frac{1}{2\pi} \Big[ \phi(\beta) - \phi(0) \Big],
\end{align}
where $W$ is the integer winding number of the phase field. In other words, $\psi'(\tau,x)$ is anti-periodic or periodic in $\beta$ depending on whether the winding number $W$ is even or odd. The gauge transformation further results in an effective chemical potential variation 
\begin{align}
\delta \mu = i \Big( \partial_{\tau} \phi/2 + V(\tau) \Big).
\end{align}
Fixing $\delta \mu=0$ in the saddle point approximation results in the Josephson relation $V(\tau) = - \partial_{\tau} \phi/2$ locking the charge and phase fluctuations, and we get the partition function
\begin{align}
Z &= \sum_W \int_W \mathcal{D}{\phi} \exp \Big(   i \pi N_{\rm g} W - \int d\tau \frac{\dot{\phi}^2}{4E_C}  \Big) \int \mathcal{D}\overline{\psi} \mathcal{D}{\psi}e^{-S_W},\nn
S_W &=  \sum_{\alpha, \beta = \su \sd} \int d\tau dx \overline{\psi}_{\alpha} \Big[  \partial_{\tau} \delta_{\alpha \beta}  + h_{\rm{nw}}^{\alpha \beta}(x) -\mu  \delta_{\alpha \beta}  \Big]{\psi}_{\beta}(\tau, x) + \Delta_0 \int d\tau dx \Big(  {\psi}_{\su}{\psi}_{\sd} + \overline{\psi}_{\sd}\overline{\psi}_{\su} \Big) \nn
&+ \sum_{\alpha  = \su \sd} \sum_{a=L,R} \int d\tau_2 d\tau_1 \overline{\psi}_{\alpha}(\tau_2, x_a ) \Sigma'_a(\tau_2, \tau_1)  {\psi}_{\alpha}( \tau_1, x_a)  ,
\end{align}
where we drop the prime sign for the electron fields, and we partition the path integral of the phase field into distinct winding number sectors. Note that the parity of the winding number W constrains the boundary condition for $\psi$ (thus Fourier expansion in boson or fermion Matsubara frequencies). We then want to integrate out the electron fields as they now become quadratic fields after Hubbard-Stratonovich transformations. We first decompose the complex Grassmann numbers into real (Majorana) Grassmann numbers
\begin{align}
&\psi_{a}(\tau) = \Big[ \gamma_{a1}(\tau) + i\gamma_{a2}(\tau) \Big]/\sqrt{2}, \nn
&\overline{\psi}_{a}(\tau) = \Big[ \gamma_{a1}(\tau) - i\gamma_{a2}(\tau) \Big]/\sqrt{2}, \nn
& \{ \gamma_{ai}, \gamma_{bj} \}=0,
\end{align}
where $a,b$ denotes the combined indices for space and spin, and $i,j=1,2$. Thus
\begin{align}
\int \mathcal{D}\overline{\psi} \mathcal{D}{\psi}\exp ( -S_W ) &=\int \mathcal{D}\Gamma \exp \Big \{ -\frac{1}{2} \int d\tau' d\tau \Gamma^{T}(\tau') \Big[ (\partial_{\tau} + i\tilde{ A}_{\rm{BdG}})\delta(\tau' - \tau) + \tilde{\Sigma}'(\tau',\tau)  \Big] \Gamma(\tau) \Big \} \nn
&=\text{Pf} \Big( -\tilde{G}^{-1}_{\rm{BdG}} + \tilde{\Sigma}' \Big) \nn
&= \exp \Big \{ \log \text{Pf}\Big( -\tilde{G}^{-1}_{\rm{BdG}} \Big) + \frac{1}{2} \log \text{det} \Big( 1 - \tilde{G}_{\rm{BdG}} \tilde{\Sigma}' \Big) \} \nn
&= Z^{\rm BdG}_{W} \exp \Big \{ \frac{1}{2} \rm{Tr} \log \Big( 1- G_{\rm{BdG}} \Sigma' \Big) \Big \},
\end{align}
where real and anti-symmetric $\tilde{ A}_{\rm{BdG}}$ is the mean-field Bogoliubov-de Gennes (BdG) Hamiltonian for the semiconductor-superconductor nanowire in the Majorana basis, $\tilde{G}_{\rm{BdG}} = - (\partial_{\tau} + i\tilde{ A}_{\rm{BdG}})^{-1}$ is the corresponding Green's function. Note that terms with tilde signs on the top are in the Majorana basis, while those without tilde signs are in the fermion basis. $Z^{\rm BdG}_{W}$ is the partition function for the isolated superconducting ring at the mean-field level, which can take only two different values depending on the parity of the winding number $W$. For even winding number $W$, 
\begin{align}
Z^{\rm BdG}_{{\rm even}~W} &=\int_{\Gamma(\beta) = - \Gamma(0) } \mathcal{D}\Gamma \exp \Big \{ -\frac{1}{2} \int d\tau \Gamma^{T}(\tau) (\partial_{\tau} + i\tilde{ A}_{\rm{BdG}}) \Gamma(\tau) \Big \} = \prod_{m} \Big( 1 + e^{-\beta \epsilon_m} \Big) \nn
& \approx   \prod_{\epsilon > 0} 2 \cosh \Big(\frac{\beta  \epsilon}{2} \Big)
\end{align}
with $\epsilon_m > 0$ being the single-particle excitation energies of the semiconductor-superconductor system. In the last step, we take out an overall factor $\prod_{\epsilon > 0} e^{\beta \epsilon /2}$ for the partition function. For odd winding number $W$,
\begin{align}
Z^{\rm BdG}_{{\rm odd}~W} &=\int_{\Gamma(\beta) = \Gamma(0) } \mathcal{D}\Gamma \exp \Big \{ -\frac{1}{2} \int d\tau \Gamma^{T}(\tau) (\partial_{\tau} + i\tilde{ A}_{\rm{BdG}}) \Gamma(\tau) \Big \} = \rm{sgn}~\rm{Pf}(\tilde{A}_{\rm{BdG}}) \prod_{m} \Big( 1 - e^{-\beta \epsilon_m} \Big) \nn
& \approx  \rm{sgn}~\rm{Pf}(\tilde{A}_{\rm{BdG}}) \prod_{\epsilon > 0} 2 \sinh \Big(\frac{\beta  \epsilon}{2} \Big),
\end{align}
where the additional factor $\rm{sgn}~\rm{Pf}(\tilde{A}_{\rm{BdG}})$ is due to the existence of a self-adjoint zero-frequency mode $\Gamma(\omega_n=0)$ in the odd winding number sector. Therefore, the total partition function for the NSN junction can be written as
\begin{align}
&Z = \sum_W Z^{\rm{BdG}}_W \int \mathcal{D} \phi \exp \Big( - S_W[\phi] \Big), \nn
& S_W[\phi] =  S^0_W[\phi] + S^{\rm leads}_W[\phi], \nn
& S^0_W[\phi] =  \int d\tau \frac{\dot{\phi}^2}{4E_C} -i \pi N_{\rm g} W, \nn
&  S^{\rm leads}_W[\phi] =  -  \frac{1}{2} \rm{Tr} \log \Big( 1- G_{\rm{BdG}} \Sigma' \Big) 
\end{align}

\subsection{IV. Generalized dissipative quantum rotor model }
We perturbatively expand the action of the leads in terms of the weak island-lead coupling:
\begin{align}
S^{\rm leads}_W[\phi] =  -  \frac{1}{2} \rm{Tr} \log \Big( 1- G_{\rm{BdG}} \Sigma' \Big) \approx \frac{1}{2} \rm{Tr} \Big( G_{\rm{BdG}} \Sigma' \Big) + \frac{1}{4} \rm{Tr} \Big( G_{\rm{BdG}} \Sigma' G_{\rm{BdG}} \Sigma' \Big),
\end{align}
where in the Nambu basis, the lead self-energy is
\begin{align}
&\Sigma'_a(\tau_1, \tau_2) = \Sigma_a(\tau_1- \tau_2) e^{i\left[ \phi(\tau_1) - \phi(\tau_2) \right]\sigma_z/2}, \nn
&\Sigma_a(\tau) =  \frac{-\Gamma_a/\beta}{\sin(\pi \tau/\beta)}.
\end{align}
Assuming that the SC gap is large so that the Green's function is local in time and space, then
\begin{align}
&\rm{Tr}( G_{\rm{BdG}} \Sigma' ) = \sum_{a=L,R} \rm{Tr}_{\sigma} \int d\tau_1 d\tau_2 G_{\rm{BdG}}(\tau_1 - \tau_2) \Sigma'_a(\tau_2, \tau_1) \propto  G_{\rm{BdG}}(0) \Sigma(0) \approx 0, \nn
& \rm{Tr}( G_{\rm{BdG}} \Sigma' G_{\rm{BdG}} \Sigma' ) \approx  \sum_{a=L,R} \int d\tau_1 d\tau_2 \rm{Tr}_{\sigma} \Big( G_{\rm{BdG}} (0; x_a ) \Sigma'_a( \tau_1, \tau_2) G_{\rm{BdG}} ( 0; x_a) \Sigma'_a( \tau_2, \tau_1) \Big),
\end{align}
where $x_{\rm L(R)}$ denotes the left (right) end of the SC island coupling to the lef (right) lead. For the lead self-energy
\begin{align}
& \Sigma'_a( \tau_1, \tau_2) = \Sigma_a( \tau_1- \tau_2)e^{i\left( \phi_1- \phi_2 \right)\sigma_z/2} =  \Sigma_a(\tau_1- \tau_2) \left( e^{i\left( \phi_1- \phi_2 \right)/2}\sigma_p + e^{-i\left( \phi_1- \phi_2 \right)/2}\sigma_h \right), \nn
& \sigma_p = 
\begin{pmatrix}
1 & 0 \\
0 & 0
\end{pmatrix},
\quad \sigma_h = 
\begin{pmatrix}
0 & 0 \\
0 & 1
\end{pmatrix}.
\end{align}
Dropping the phase-independent product, we get
\begin{align}
S^{\rm leads}_W[\phi]  &\approx \frac{1}{4} \rm{Tr} \Big( G_{\rm{BdG}} \Sigma' G_{\rm{BdG}} \Sigma' \Big) \nn
&  \approx \frac{1}{4} \sum_a \rm{Tr}_{\sigma} \Big( G_{\rm{BdG}}(x_a) \Gamma_a \sigma_p G_{\rm{BdG}}(x_a) \Gamma_a \sigma_h \Big) \int \frac{d\tau_1 d\tau_2}{\beta^2}  [\Sigma(\tau_1- \tau_2)/\Gamma_a ]^2 \cos(\phi(\tau_1) - \phi(\tau_2)) \nn
&= \frac{(g_L + g_R)}{2}  \int \frac{d\tau_1 d\tau_2}{\beta^2}  \frac{ \cos( \phi(\tau_1) - \phi(\tau_2) )}{\sin^2(\pi (\tau_1 - \tau_2)/\beta)}
\end{align}
where $g_a = \frac{1}{2}\rm{Tr}_{\sigma} \Big( G_{\rm{BdG}}(x_a) \Gamma_a \sigma_p G_{\rm{BdG}}(x_a) \Gamma_a \sigma_h \Big) $ is the local Andreev conductance between the lead and the SC island. Therefore we have obtained the action for the generalized AES model~\cite{Ambegaokar1982Quantum} which includes the ground state parity of the SC island:
\begin{align}
&Z = \sum_W Z_W =  \sum_W Z^{\rm{BdG}}_W \int \mathcal{D} \phi \exp \Big( - S_W[\phi] \Big), \nn
& S_W[\phi] = \int d\tau \frac{\dot{\phi}^2}{4E_C} -i \pi N_{\rm g} W+ g_0\int \frac{d\tau_1 d\tau_2}{\beta^2}  \frac{1- \cos( \phi(\tau_1) - \phi(\tau_2) )}{\sin^2(\pi (\tau_1 - \tau_2)/\beta)},
\label{eq:effectiveAction}
\end{align}
where $g_0 = (g_L + g_R)/2$ is the averaged conductance denoting the tunneling to the external leads.


\subsection{V. Conductance for the dissipative quantum rotor model}
The conductance for the NSN junction at zero-bias voltage can be expressed as the correlation function~\cite{Bascones2000Nonequilibrium}
\begin{align}
G = g_0 \mathcal{G}( \beta/2)= \Big \langle e^{ i \phi(\beta/2) - i \phi(0) } \Big \rangle.
\end{align}
Like the partition function, any equilibrium observable can be expanded in $W$ sectors and evaluated independently in each, i.e., the correlation function can be expressed as
\begin{align}
&\mathcal{G}( \beta/2)=\mathcal{G}_0 + \Big( \mathcal{G}_1 -\mathcal{G}_0 \Big) \frac{Z_1}{Z_0}  + \Big( \mathcal{G}_{-1} -\mathcal{G}_0 \Big) \frac{Z_{-1}}{Z_0} + ..., \nn
&\mathcal{G}_W = \Big \langle e^{ i \phi(\beta/2) - i \phi(0) } \Big \rangle_W =  \frac{ \int \mathcal{D} \phi e^{ i \phi(\beta/2) - i \phi(0) }  \exp \Big( - S_W[\phi] \Big) }{ \int \mathcal{D} \phi \exp \Big( - S_W[\phi] \Big) }
\end{align}
where $Z_W$ is defined in Eq.~\eqref{eq:effectiveAction}, and we ignore the sectors of larger winding numbers. In each winding number sector $W$, the phase field can be expanded around the classical trajectory
\begin{align}
& \phi(\tau) = 2\pi W \tau / \beta + \delta \phi(\tau), \nn
\end{align}
with $\delta \phi(\beta) = \delta \phi(0)$. To the zeroth order in $\delta \phi(\tau)$, $\phi(\tau) = 2\pi W \tau / \beta$ so that 
\begin{align}
Z^{(0)}_W = Z^{\rm{BdG}}_W \int \mathcal{D} \phi \exp \Big( - S^{(0)}_W[\phi] \Big)
= Z^{\rm{BdG}}_W \exp \Big( - \frac{\pi^2W^2}{ \beta E_C} + i \pi N_{\rm g} W - 2g_0 |W| \Big),
\end{align}
since $\int^1_0 dx \frac{1-\cos(2\pi W x)}{\sin^2(\pi x)} = 2|W|$. The ratio of partition function between winding number one and number zero will be 
\begin{align}
Z_{\pm 1}/ Z_0 \approx Z^{(0)}_{\pm 1}/ Z^{(0)}_0 = \Big( {\rm sgn}~{\rm Pf }\tilde{A}_{{\rm BdG}} \Big) \prod_{\epsilon > 0} 2 \tanh \Big(\frac{\beta  \epsilon}{2} \Big) \exp ( \pm i \pi N_{{\rm g}} )  \exp \Big( - \frac{\pi^2}{ \beta E_C}  - 2g_0 \Big).
\end{align}
For the correlation function at zeroth order, $\mathcal{G}^{(0)}_W = (-1)^W$. If the phase fluctuations are included up to the quadratic order, 
\begin{align}
\mathcal{G}_W = (-1)^W \prod_{n=1,3,5,...} \exp \Big( - \frac{1}{ \frac{\pi^2 n^2}{2\beta E_C}  + g_0( n - |W| ) \Theta(n - |W|) } \Big),
\end{align}
where $\Theta(x)$ is the heaviside step function. The additional factor due to the quadratic phase fluctuations depends weakly on the averaged conductance $g_0$ in the weak Coulomb blockade limit $\beta E_C \ll 1$.


\end{document}